\documentclass[aps,
prd,
preprint,
nofootinbib,
groupedaddress]{revtex4}

\usepackage{amsmath,amssymb}
\usepackage{mathrsfs,graphicx}

\usepackage{pstricks,pst-node,pst-coil}

\makeatletter
  \def\ifundefined#1{\expandafter\ifx\csname#1\endcsname\relax}
  \ifundefined{default@color}
    \let\default@color=\current@color
  \fi
\makeatother

\newcommand{\D}{\Delta}

\newcommand{\p}{\partial}
\renewcommand{\t}{\theta}
\renewcommand{\o}{\omega}

\newcommand{\Dr}{\frac{\D}{r}}
\newcommand{\rD}{\frac{r}{\D}}

\newcommand{\half}{\frac{1}{2}}
\newcommand{\eps}{\varepsilon}

\newcommand{\beq}{\begin{equation}}
\newcommand{\eeq}{\end{equation}}
\newcommand{\bea}{\begin{eqnarray}}
\newcommand{\beas}{\begin{eqnarray*}}
\newcommand{\mO}{\mathcal{O}}
\newcommand{\eea}{\end{eqnarray}}
\newcommand{\eeas}{\end{eqnarray*}}
\newcommand{\bD}{\Delta}
\newcommand{\dw}{\delta\omega}
\begin{document}
\title{Compact extra-dimensions as solution to the strong CP problem}
\author{F.L. Bezrukov}
\email{Fedor.Bezrukov@mpi-hd.mpg.de}
\affiliation{
  Max-Planck-Institut f\"ur Kernphysik,\\
  PO Box 103980, 69029 Heidelberg, Germany\\
  Institute for Nuclear Research of Russian Academy of Sciences,\\
  Prospect 60-letiya Oktyabrya 7a, Moscow 117312, Russia
}
\author{Y. Burnier}
\email{yburnier@physik.uni-bielefeld.de}
\affiliation{Faculty of Physics, University of Bielefeld\\
D-33501 Bielefeld, Germany
}
\begin{abstract}

We propose a way of a dynamical solution of the strong CP in models with compact extra-dimensions.  To this aim we consider a one dimensional toy model for QCD, which contains a vacuum angle and a strong CP like problem.  We further consider a higher dimensional theory, which has a trivial vacuum structure and which reproduces the perturbative properties of the toy model in the low-energy limit.  In the weak coupling regime, where our computations are valid, we show that the vacuum structure of the low-energy action is still trivial and the strong CP problem does not arise.  Also, no axion-like particles are generated in this setup.
\end{abstract}
\pacs{11.10.Kk, 11.27.+d}

\preprint{BI-TP 2008/34; arXiv:0811.1163}
\maketitle 

\section{Introduction}
Quantum Chromodynamics has a complicated vacuum structure \cite{Jackiw:1976pf}. Due to the nontrivial group of mappings from the coordinate space to the gauge group $\pi^3(SU(3))=\mathbb{Z}$, there is an infinite number of perturbative vacua $|n\rangle$ which are not connected by smooth gauge transformations. Physical passage between these vacua is possible at an exponentially suppressed rate. One can further define physical vacua, that is to say, vacua between which transition is not possible. These are called $\theta$-vacua and are expressed as discrete Fourier transform of the $n$-vacua, $|\theta\rangle=\sum_{n=-\infty}^\infty e^{in\theta}|n\rangle$. The physics depends on which vacua we are in, so $\theta$ becomes a new parameter of the theory. Equivalently this may be expressed by a new effective term in the Lagrangian of the form $i\theta q(x)$, where $q(x)$ is the topological charge density.\footnote{$q(x)=\frac{1}{32\pi^2}\tilde{F}^{\mu\nu}F_{\mu\nu}(x)$.} The presence of this term implies that $\langle q(x)\rangle$ does not vanish and is an odd function of $\theta$ \cite{Luscher:1978rn}. The non-vanishing of $\langle q(x)\rangle$ leads in turn to the CP-violation and can be measured experimentally. In principle $\theta$ can take any value in the interval $[0,2\pi)$, but according to the measurement of the neutron dipole moment, it turns out to be extremely small $\theta<2\cdot10^{-10}$. The fact that this value is unnaturally small is known as the strong CP problem.

Different models were proposed to explain the smallness of $\theta$. Some scenarios \cite{Dvali:2005zk} lead to a larger probability of living in a universe with small $\theta$, but most of the models rely on the addition of some dynamical properties to $\theta$.
A possible way is to promote the parameter $\theta$ to a field, the axion, which will relax dynamically to zero \cite{Peccei:1977hh}. This solution is however, in its minimal setup, ruled out by the non-detection of this axion particle \cite{Asztalos:2006kz}. More complicated models exist and can avoid these bounds, some of these also use extra-dimensions and allow the axion to propagate in the bulk \cite{Dienes:1999gw}. 

Though both the CP problem and its axion solution have their origin in the symmetry properties of the system (existence of large gauge transformation for the first, and Peccei-Quinn symmetry for the second), one should be careful with the high energy extensions of the theory.  In case where a high energy effect break some of the symmetries, the situation may change.   This, for example, happens in the case of the global Peccei-Quinn symmetry.  It is usually considered that it should be broken by higher order operators arising from quantum gravity corrections\cite{Barr:1992qq}. The contributions from these operators scale like a power of $f_a/M_{Pl}$, where $f_a$ is the axion coupling constant. 
Due to the astrophysical bounds from supernova cooling by light particles, $f_a$ has to be large and the ratio $f_a/M_{Pl}$ cannot be small enough, leading to corrections to the potential that spoils the solution of the strong CP problem. These problems are avoided in invisible axion models\cite{Kim:1979if}, where the Peccei-Quinn symmetry is a consequence of gauge invariance, and thus survives the quantum gravity effects.
If the high energy theory does not violate the Peccei-Quinn symmetry, the solution to the CP problem is unaffected.  On the other hand, if the high energy theory breaks the invariance against large gauge transformation, this may remove the CP problem by itself.  This is the essence of the solution of the CP problem proposed in this article.

Extra dimensions can provide a solution to the strong CP problem without light axion particles \cite{Khlebnikov:1987zg,Chaichian:2001nx,Khlebnikov:2004am}. As we will see, the solution in this setup is also dynamical, but does not imply the existence of some light particle or any additional degree of freedom.  The basic idea is that the group of mappings of a higher dimensional space to the gauge group ($\pi^d(SU(3)), d\leq 4$) is trivial\footnote{Except $\pi^4(SU(3))=\mathbb{Z}_2$, which should not be an issue for our discussion.} and $\theta$ vacua cannot be defined. This type of solutions were already proposed some time ago with compact extra dimensions \cite{Khlebnikov:1987zg} or with infinite ones \cite{Chaichian:2001nx}. It was however pointed out in \cite{Khlebnikov:2004am} that the solution might be more complicated than expected from the simple topological argument. The presence of a suitable extra dimension removes the degeneracy of the $|n\rangle$ vacua, thus preventing to define $\theta$-vacua. This solution is only formal and the strong CP problem may reappear when we require that the observable effects of the extra dimensions are small. In this limit it can be expected that the usual QCD is recovered and the strong CP problem reappear in some way.\footnote{This is indeed the case in \cite{Khlebnikov:2004am} when the extra-dimension is taken to be an orbifold.}
To show that the strong CP problem actually disappear, we should not only get a trivial vacuum structure, but also show that the physical consequences of a non-vanishing $\theta$ also do disappear, that is to say, the topological charge should vanish in average ($\langle q\rangle=0$).

To our best knowledge, no complete setup where fields are explicitly localised
on a brane and where the strong CP problem is solved have been constructed
till now. Solving the strong CP problem in real QCD using extra dimensions is
difficult. The full QCD is complicated and the localisation of non-Abelian
theories on a brane is non-trivial \cite{nonAlocal}. In particular the gauge
theories in more than 4 dimensions are generically not renormalisable
(while $SU(N)$ gauge theories might however be nonperturbatively
renormalisable in
5 or even 6 dimensions, see \cite{Gies:2003ic}).
Thus, an eventual
extra-dimensional extension of QCD might need an UV completion, that is to
say will itself have to be thought as an effective theory comming as a low
energy limit of some renormalisable theory (for instance string theory).
However, certain important features the solution of the strong CP problem can
be addressed in a simplified model. We will consider here the simplest
possible model where a strong CP like problem occurs and can be solved by the
addition of an extra dimension. Keeping this model as simple as possible will
enable us to perform the relevant computations explicitly, retaining all the
important topological properties.

The simplest model which contains a QCD like $\theta$ angle is the Abelian Higgs model in 1+1 dimensions. This model has the same complicated vacuum structure as QCD ($\pi^1(U(1))=\mathbb{Z}$) and its effective action also contains a $\theta$-parameter. This model was already studied many times as a toy model for QCD. For instance the QCD $U(1)$ problem was first solved there \cite{Kogut:1974kt}. This solution could then be mapped to QCD \cite{tHooft}.
Note that the Abelian Higgs model was also successfully used as a toy model for the electroweak baryogenesis \cite{AHMEW,Burnier:2005he}. Our calculations could also be relevant to such issues.

We then consider a higher dimensional theory, 
which reproduce the perturbative properties of the Abelian Higgs model as a low-energy action.
In this latter theory, the spatial dimensions have the topology of a sphere, so that the vacuum structure is trivial ($\pi^2(U(1))=0$). At low energy, particles are localised on the equator (which will be the brane) and the extra dimension extends towards the poles. To simplify the computations we use in practice a geometrically simpler realization of this topology, and consider a pancake (the sphere is flattened to two disks).

This article is structured as follows. We first review the relevant properties of the Abelian Higgs model in 1+1 dimensions in Section 2. In Section 3, we consider the Abelian Higgs model in 2+1 dimensions and localise the fields with the help of a warp factor. We show that the 1+1 dimensional theory is correctly recovered as low energy effective action and discuss the validity of the classical Kaluza-Klein decomposition in some detail. In Section 4, we discuss the non-linear sector of the theory. The 2+1 dimensional equivalent of the 1+1 dimensional vacuum structure is presented. We show that the degeneracy of the 1+1 dimensional $|n\rangle$ vacua is lifted and this new structure is interpreted as a potential for the topological charge density. Finally the evolution of the topological charge density is considered. At least in the weak coupling regime where our calculations are valid, it relaxes very fast to zero, which solves the strong CP problem. We conclude in Section 6 and discuss the extension to more realistic models.

\section{Toy model}

Toy models to address the strong CP problem in the framework of extra-dimensions were already considered. In \cite{Khlebnikov:2004am}, 2+1 dimensional electrodynamics was localised on a one dimensional brane. When two spatial dimensions form a sphere, the topological charge density $\langle q(x)\rangle$ is shown to get the effective action of a harmonic oscillator with a small frequency. This kind of dynamics does not completely solve the $\theta$ problem since $\langle q(x)\rangle$ is slowly oscillating in time. Furthermore, electrodynamics in 1+1 dimensions is not really adapted to our purposes. It a is trivial theory and contains a strong CP like problem only when charged particles are added. In \cite{Khlebnikov:2006yq} the effective low dimensional action from \cite{Khlebnikov:2004am} was reconsidered with charged fermions and the strong CP problem was addressed but without reference to some extra-dimensional scenario.
A simple model that contains $\theta$-vacua and a strong CP like problem is the Abelian Higgs model in 1+1 dimensions. This is the model that we will consider and before discussing its embedding into a higher dimensional space, we recall some of its properties \cite{AHM}.

\subsection{Basic properties of the Abelian Higgs model in 1+1 dimensions}

The action for the Abelian Higgs model in 1+1 dimensions reads:
\beq
S =\int dxdt\left( -\frac{1}{4}F_{\mu \nu }F^{\mu \nu }-V(\phi )+\frac{1}{2}\left| D_{\mu }\phi \right|
^{2}\right),\label{Lmod1}
\eeq
where $D_{\mu}=\partial _{\mu}-i\tilde{e}A_{\mu}$, and $\phi=(H+\tilde{v}) e^{i\sigma}$ is a complex scalar field with the symmetry breaking potential
\begin{equation}
V(\phi)=\frac{\tilde{\lambda}}{4}\left(|\phi|^2-\tilde{v}^2\right)^2.\label{V(phi)}
\end{equation}
Note that we use tilde everywhere to denote 1+1 dimensional variables.
We will now discuss the perturbative properties of this action. We want to identify the degrees of freedom, derive their spectrum and interactions.
As we will latter compare the effective action of the 2+1 dimensional Abelian Higgs model localised on the boundary of the disk to the 1+1 dimensional Abelian Higgs model, we will directly consider the space dimension to be a circle of length $2\pi R$. We always consider $1/R$ to be much smaller that the particle masses. 
To extract the physical degrees of freedom in the simplest way, we use the unitary gauge ($\sigma=0$). 
We replace the coordinate $x\to R\theta$ and decompose the fields in partial waves
\bea
A_\mu&=&\frac{1}{\sqrt{2\pi R}}\sum_n e^{in\t}A_\mu^n,\notag\\
H&=&\frac{1}{\sqrt{2\pi R}}\sum_n e^{in\t}H^n,\label{pw}
\eea
with $(A^n)^*=A^{-n}$ and $(H^n)^*=H^{-n}$, so that the fields $A_\mu,H$ are real.
Substituting the partial wave expansion (\ref{pw}) into the action (\ref{Lmod1}) we get
\bea
S=\int dt \sum_n \left[  \half|\dot H^n|^2-\half\left(2\tilde{\lambda}\tilde{v}^2+\frac{n^2}{R^2}\right)|H^n|^2-\frac{\tilde{\lambda}\notag \tilde{v}}{\sqrt{2\pi R}}\sum_mH^n  H^mH^{-n-m}\right.\\\left.-\frac{\tilde{\lambda}}{8\pi R}\sum_{m,k}H^nH^mH^kH^{-n-m-k}\right]+S_{GF}.\label{SH11}
\eea
We see that after symmetry breaking, the physical spectrum contains a scalar field (Higgs) with the mass $m_H^2=2\tilde{\lambda} \tilde{v}^2$ and its dispersion relation is, as expected, $\o_n^2=m_H^2+n^2/R^2$. The gauge field action reads
\bea
S_{GF} &=&\int dt \sum_n \left[ \half|\dot A_1^n|^2-\frac{\tilde{e}^2\tilde{v}^2}{2}|A_1^n|^2+\half\left(\tilde{e}^2\tilde{v}^2+\frac{n^2}{R^2}\right)|A_0^n|^2\right.\notag\\
&&+\frac{in}{2R}\left(\dot A_1^n A_0^{-n}-\dot A_1^{-n} A_0^n\right)+\frac{\tilde{v} \tilde{e}^2}{\sqrt{2\pi R}}\sum_m (A_0^nA_0^m-A_1^nA_1^m)h^{-n-m}\notag\\ 
&&\left.+\frac{\tilde{e}^2}{4\pi R}\sum_{m,k}(A_0^nA_0^k-A_1^nA_1^k)h^mh^{-n-m-k}\right].\label{SG11}
\eea
The field $A_0$ is not dynamical and can be integrated out using its equations of motion,
\beq
A_0^2\left(\tilde{e}^2\tilde{v}^2+\frac{n^2}{R^2}\right)+\frac{in}{R}\dot A^n_1+\frac{2\tilde{e}^2}{\sqrt{2\pi R}}\tilde{v}\sum_k A_0^k H^{n-k}+\frac{\tilde{e}^2}{2\pi R}\sum_{m.k}A_0^k H^m H^{n-m-k}=0. \label{eoma0}
\eeq
Inserting the latter equation (\ref{eoma0}) in the gauge field action (\ref{SG11}) and rescaling 
\beq
A_1^n\to\sqrt{\frac{\tilde{e}^2\tilde{v}^2+n^2/R^2}{\tilde{e}^2\tilde{v}^2}}A_1^n
\eeq
we are left with (up to the terms $\mO((\tilde{e}\tilde{v}R)^{-2})$)
\bea
S_{GF} =\int dt \sum_n \left[ \half|\dot A_1^n|^2-\half\left(\tilde{e}^2\tilde{v}^2+\frac{n^2}{R^2}\right)|A_1^n|^2-\frac{\tilde{v} \tilde{e}^2}{\sqrt{2\pi R}}\sum_m A_1^nA_1^mh^{-n-m}\right.\notag\\\left.-\frac{\tilde{e}^2}{4\pi R}\sum_{m,k} A_1^nA_1^k h^mh^{-n-m-k}+ \textrm{higher order interactions}\right].\label{9}
\eea
This is the action of a scalar field with the mass $m_W=\tilde{e}\tilde{v}$. Integrating out $A_0$ leads to some complications, such as $\mO((\tilde{e}\tilde{v}R)^{-2})$ corrections in the interactions terms and further higher order (in the fields) interactions. In the following we will mainly perform semi-classical calculations. To insure that they are valid, we will consider only small couplings. Consequently, corrections in (\ref{9}) are small and not relevant to our discussion.\footnote{They are not needed to compare with the effective action for localised 2+1 dimensional fields since we will make similar approximations there.}
Note that the couplings in the Lagrangian (\ref{Lmod1}) are dimensionfull. Weak coupling criterion means that
\beq
\frac{\tilde{\lambda}}{m_H^2}\sim \frac{\tilde{e}^2}{m_W^2}\ll 1\quad\Leftrightarrow\quad \tilde{v}\gg 1 \label{weakcoupling11}
\eeq
Of course the QCD coupling is large and this toy model is not realistic in this sense. However the strong CP problem does not rely on the fact that QCD is strongly coupled and we may very well use a simple toy model which enables simple perturbative calculations.

As emphasised in the introduction, the Abelian Higgs model has almost identical non-perturbative properties as QCD. We list them in the following and, when it happens, point out where they differ from QCD.

\subsection{Vacuum structure}

In short, the ensemble of vacuum configurations are the ensemble of mappings from the space $S^1$ to the gauge group $U(1)$. This ensemble can be divided into equivalence classes identifying vacua that can be related by a smooth gauge configuration. Mathematically this is represented by the homotopy group $\pi^1(U(1))=\mathbb{Z}$. This means that there is a discrete infinity of classes that cannot be related by a smooth gauge configuration. 
An element of the class $n$ is for instance
\beq
\phi^{(n)}=\tilde{v}\exp\left(\frac{ i n x}{R}\right), \quad A^{(n)}_1=\frac{ n}{\tilde{e}R},\quad A_0=0.\label{nvacua}
\eeq
These equivalence classes can be distinguished by the Chern-Simons number:
\beq
N_{CS}=\frac{\tilde{e}}{2\pi}\int dx\;A_1(x).\label{ncs}
\eeq
It takes the value $n$ when applied on a vacuum state of the equivalence class $n$. 
The transition from one equivalence class to another can formally be achieved by a discontinuous gauge transformation. An example of the gauge transformation that changes the Chern-Simons number by $n$ is
\beq U^n=\exp\left(\frac{ i n x}{R}\right).\eeq

\subsection{Sphaleron}

\begin{figure} 
\begin{center} 
\includegraphics[width=120mm,height=35mm]{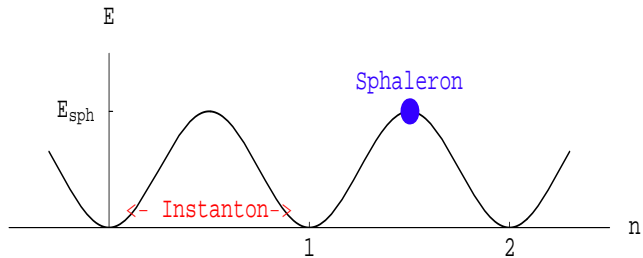} 
\caption{Sketch of the vacuum energy as function of the winding number $n$}
\label{pot2D}
\end{center}
\end{figure}

Physically, the transition between two vacua needs to go through a set of non-vacuum configurations that form an energy barrier (see Fig.\ \ref{pot2D}). For instance, the set of static field configurations
\begin{eqnarray}
\phi ^{cl} &=&\tilde{v}e^{\frac{ ix\tau }{R}}\left[ \cos (\pi \tau )+i\sin (\pi
\tau )\tanh (m_H x\sin (\pi \tau ))\right], \notag  \\
A_{1}^{cl} &=&\frac{\tau }{\tilde{e}R}  \label{2config},
\end{eqnarray}
form a path that goes from vacuum $n=0$ at $\tau=0$ to vacuum $n=1$ at $\tau=1$, minimising the energy of the intermediate configurations. The configuration of maximal energy 
\beq E_{sph}=\frac{2}{3}m_H \tilde{v}^2\label{Esph}\eeq
is reached at $\tau=\half$, and is called the  sphaleron. It is relevant for the high temperature behaviour of the theory. Thermal fluctuations can reach the required energy $E_{sph}$ and the system may pass classically between vacua.
Although all the previous and further discussed properties of the Abelian Higgs model exactly match QCD, we should point that, strictly speaking, the sphaleron does not exist in QCD. The corresponding configurations in QCD contain a free scale parameter and their potential energy can be made infinitely small in the limit of large configuration size. However, at large scale, kinetic energy along the path grows and the transition rate is still suppressed. This small difference will not be of importance for our discussion.

\subsection{Instanton}

At small or vanishing temperature, the system can also tunnel from one vacuum to another. In quantum field theory, tunnelling is represented by instantons, which are solutions of the classical equations of motion in Euclidean space-time. The set of configurations (\ref{2config}) can serve as Ansatz to compute the Euclidean action, which gives the leading information to compute the tunnelling rate.

The proper instanton in this model, i.e.\ the field configuration that minimises the Euclidean action and describes the tunnelling between  the states $|0\rangle$ and $|n\rangle$ is the Nielsen-Olesen vortex \cite{Nielsen:1973cs} with 
winding number 
\beq
Q=\Delta N_{CS}=\frac{\tilde{e}}{4\pi}\int \eps_{\mu\nu}F^{\mu\nu}d^2x=n.
\eeq
Parametrising the Euclidean space-time in polar coordinates $(r,~\theta )$ the field configuration reads:  
\begin{eqnarray} 
\phi (r,\theta ) &=&e^{in\theta }f(r)  
\label{phi}, \\ A^{i}(r,\theta ) &=&\varepsilon 
^{ij}\widehat{r}^{j}A(r)   
\label{AA}, 
\end{eqnarray}
where $\widehat{r}$ is the unit 
vector $\widehat{r}=(\cos\theta,\ \sin\theta)$ and $\varepsilon 
^{ij}$ the completely antisymmetric tensor with $\varepsilon 
^{01}=1$.
The functions $A$ and $f$ have to satisfy the 
following limits:
\begin{eqnarray} 
&&f(r)\overset{r\rightarrow 0}{\longrightarrow }c r^{\left| n\right| } 
\label{l1},\notag\\ 
&&f(r)\overset{r\rightarrow \infty }{\longrightarrow }\tilde{v}, 
\label{l2}\notag \\ 
&& A(r)\overset{r\rightarrow 0}{\longrightarrow }0, 
\label{l3}\\ 
&&A(r)\overset{r\rightarrow \infty }{\longrightarrow }-\frac{n}{\tilde{e}r}. 
\label{l4} \notag
\end{eqnarray} 
Introducing dimensionless variables 
\beq
r\to\frac{r}{\sqrt{\lambda} v},\quad A\to\frac{\sqrt{\lambda} v}{e} A,\quad f\to\frac{\sqrt{\lambda} v}{e}f,\label{dimless}
\eeq 
the classical action of the instanton is
\bea S_{cl}&=&\pi \tilde{v}^2\mu^2 \int_0^\infty 
{r}d{r}\left\{\left({A}'(r)+\frac{{A}(r)}{{r}}
\right)^2+{f}'(r)^2+ 
{f}(r)^2\left({A}(r)-\frac{1}{{r}}\right)^2\right.\notag\\ &&+\left.
\frac{\mu^2}{2} 
\left({f}^2(r)-\frac{1}{\mu^{2}}\right)^2\right\}= \pi \tilde{v}^2 \mu^2 B(\mu)\label{Evor1},
\eea
where $\mu^2=\tilde{\lambda}/\tilde{e}^2$ and the function $B(\mu)$ is of order one, and depends weakly on its argument.

\subsection{$\theta$-vacua and topological charge}
%
We shall consider the non-local gauge transformation $U^1$ that changes the Chern-Simons number by one. It commutes with the Hamiltonian and is unitary. The operator $U^1$ can therefore be diagonalised simultaneously with the Hamiltonian and must have eigenvalues of modulus 1. The $\theta$-vacua are defined as a superpositions of the $|n\rangle$ states which are at the same time an eigenvector of $U^1$ with eigenvalue $e^{i\theta}$:
\beq
|\theta\rangle=\sum_n e^{in\theta}|n\rangle.\label{thetavacua}
\eeq
A physical transition between the $\theta$-vacua is not possible. They form different sectors with different physical properties and $\theta$ is a parameter of the theory.
The topological charge density $q=\eps_{\mu\nu}F^{\mu\nu}$ in a $\theta$-vacuum does not vanish in average \cite{Luscher:1978rn},
\beq
\langle q \rangle=8\pi K e^{-S_{cl}} \sin\theta,
\eeq
with $S_{cl}$ the instanton action and the factor $K\sim m_H$ takes quantum corrections into account.

\section{2+1 dimensional Abelian Higgs model on a disk}

We will now recreate the 1+1 dimensional Abelian Higgs model as the low energy effective theory for a 2+1 dimensional model. The perturbative properties of our low energy theory has to resemble closely the 1+1 dimensional Abelian Higgs model, but hopefully have a different vacuum structure. We expect that this is possible since, as $\pi^2(U(1))=0$, no $\theta$ vacua exists in the original 2+1 dimensional theory. 

As discussed in the introduction, we will suppose that two spatial dimensions are the surface of a pancake, where the low energy fields live on its boundary and the extra-dimension extends from the boundary towards the centre. This pancake can be adroitly sliced into two disks. By symmetry we can expand the fields into functions that are odd or even with respect to pancake flipping. In the following we will only consider one disk where the fields can have either Neumann or Dirichlet boundary conditions on its boundary.

In 2+1 dimensions, we consider a complex Higgs field $\phi$ and an Abelian gauge field. The space time is a disk, which we parametrise with polar coordinates $(t,\theta,r)$. We will also introduce a warp factor $\Delta(r)$ to localise the fields \cite{nonAlocal,warpedgf}. Note that we do not want to consider gravity here and the warp factor will be thought as coming from the coupling to some external classical field, which multiplies the whole action,
\begin{equation}
S=\int dr~d\theta~ dt \sqrt{g} \bD(r)\left[-\frac{1}{4} g^{AB} g^{CD} F_{AC} F_{BD}+ \half g^{AB} (D_A\phi)^*(D_B\phi)-V(\phi)\right]\label{LH},
\end{equation}
with the Higgs potential $V(\phi)=\frac{\lambda}{4}\left(|\phi|^2-v^2\right)^2$ and $D_M=\partial_M-ieA_M$. The metric of the disk is
\begin{equation*}
        g_{MN}= 
        \begin{pmatrix}
                1 & & \\
                & -r^2 & \\
                & & -1
        \end{pmatrix},
\end{equation*}
and the volume element $\sqrt{g}=r$.
The action can be rewritten more explicitly as
\bea
S&=\int dr~d\theta~ dt\bD &\left\{\frac{1}{2}  
\left[ \frac{1}{r} F^2_{0\theta} + rF^2_{0r} - \frac{1}{r} F^2_{\theta r} \right]\right.\\&&+\left.\half \left[r|D_0\phi|^2-r|D_r\phi|^2-\frac{1}{r}|D_\t\phi|^2-r\frac{\lambda}{2}\left(|\phi|^2-v^2\right)^2\right]\right\}\notag.
\eea
In the following, we will use 
\beq \Delta(r)=e^{-2M|R-r|}.\label{D} \eeq 
The exact form of $\D$ is merely chosen to simplify the algebra. It is only needed that $\int dr \D(r)<\infty$ and that $\D(r)$ decreases from the brane ($r=R$) towards the extra dimension. 
In the following we will consider the following relation between the parameters: $M\gg m_H\sim m_W\gg 1/R$.

\subsection{Linearised theory and spectrum}

We decompose the Higgs field in polar coordinates as $\phi=(v+H)e^{i\sigma}$, and retain the quadratic part of the total (gauge field + Higgs) Lagrangian ;

\bea
\mathscr{L}_{0} &=& \frac{1}{2} \Delta(r) \notag
\left[ \frac{1}{r} F^2_{0\theta} + rF^2_{0r} - \frac{1}{r} F^2_{\theta r} \right]\\&&+\frac 1 2 \bD v^2\left[r(\p_0\sigma-e A_0)^2-r(\p_r\sigma-eA_r)^2-\frac{1}{r}(\p_\t \sigma-eA_\t)^2\right]\notag\\&&-r\bD\frac{m_H^2}{2}H^2+\frac{\bD}{2}\left[r(\p_0 H)^2-r(\p_rH)^2-\frac{1}{r}(\p_\t H)^2\right],
\eea
with $m_H^2=2\lambda v^2$. 
The linearised equations of motion read (with the rescaling $\sigma\to e \sigma$)
\bea
-\p_0(\p_0\sigma-A_0)+\frac{1}{r\D}\p_r r\D(\p_r\sigma-A_r)+\frac{1}{r^2}\p_\t(\p_\t\sigma-A_\t)&=&0,\label{Es}\\
\frac{1}{r\D}\p_r r\D(\p_0 A_r-\p_r A_0)+\frac{1}{r^2}\p_\t(\p_0 A_\t-\p_\t A_0)+ m_W^2(A_0-\p_0\sigma)&=&0,\label{EA0}\\
-\p_0(\p_0A_\t-\p_\t A_0)+\rD\p_r\Dr(\p_rA_\t-\p_\t A_r)-m_W^2(A_\t-\p_\t\sigma)&=&0,\label{EAt}\\
-\p_0(\p_0A_r-\p_rA_0)-\frac{1}{r^2}\p_\t(\p_rA_\t-\p_\t A_r)-m_W^2(A_r-\p_r\sigma)&=&0,\label{EAr}\\
\left(\frac{1}{r\bD}\p_r\bD r\p_r +\frac{1}{r^2}\p_\t^2-\p_0^2-m_H^2\right)H&=&0.\label{EH}
\eea
Note that the phase $\sigma$ and the gauge fields $A_M$ decouple from the physical Higgs $H$. These two sectors are analysed separately.

\paragraph{Higgs sector:}

After Fourier transform
\begin{equation}
        H(t, \theta, r) = \int \!\! \frac{d\omega}{2\pi} \; e^{-i\omega t} \sum_n e^{in\t} h^n(r)\label{FTh},
\end{equation} 
the Higgs field $h^n$ satisfies the following equation of motion:
\beq
\left(\frac{1}{r\bD}\p_r\bD r\p_r-\frac{n^2}{r^2}-m_H^2+\omega^2\right)h^n(r)=0. \label{equah}
\eeq
The explicit solution of the equation (\ref{equah}) which satisfy the boundary conditions $h^n(r)\propto r^n$ at $r\to0$, is given in terms of the Laguerre function:
\bea
h^n(r)\propto e^{-r (M+\Omega_H )} (r)^n L_{-\frac{M}{2 \Omega_H }-n-\frac{1}{2}}^{2 n}(2 r \Omega_H )\label{hsg},
\eea
with $ \Omega_H=\sqrt{M^2+m_H^2-\omega^2}$. We should further impose that either $h'(r)$ (symmetric mode) or $h(r)$ (antisymmetric mode) vanishes at $r=R$.
For $n=0$, the lowest mode is symmetric $h_0^0=\text{const}$ and has the energy $\omega_0=m_H$. 

For $n\neq 0$, there is a low energy symmetric mode that satisfy, in the limit $M\gg m_H$ the usual dispersion relation $\omega_0^n=\sqrt{m_H^2+\frac{n^2}{R^2}}+\mO(e^{-2MR})$ (see Appendix \ref{A}), so that we have a Higgs particle on the brane, with mass $m_H$ and, comparing to (\ref{SH11}), correct dispersion relation.

The next mode is antisymmetric and has $\omega_1\gtrsim M$. Further modes have higher energy and build a discrete spectrum (labeled with $k$) which satisfy $\omega_{k+1}-\omega_k=\frac{\pi}{2R}$ at large $k$ (see Appendix \ref{A}).

\paragraph{Gauge sector:}
After Fourier transform 
\begin{equation}
        A_{N}(t, \theta, r) = \int \!\! \frac{d\omega}{2\pi} \; e^{-i\omega t} \sum_n e^{in\t} a^n_{N}(r)\label{FTA},
\end{equation}
the equations of motion for $\sigma,a^n_0,a^n_\t,a^n_r$ are
\bea
i\omega(-i\o\sigma-a^n_0)+\frac{1}{\bD r}\p_r\bD r(\p_r\sigma-a^n_r)+\frac{in}{r^2}(in\sigma-a^n_\t)&=&0,\label{es}\\
\frac{-1}{\D r}\p_r\left(\D r(i\o a^n_r+\p_ra^n_0)\right)+\frac{1}{r^2}(n\o a^n_\t+n^2a^n_0)+(a^n_0+i\o\sigma) m_W^2&=&0,\label{ea0}\\
\o^2a^n_\t +n\o a^n_0+\rD\p_r\Dr\left(\p_ra^n_\t-ina^n_r\right)-(a^n_\t-in\sigma) m_W^2&=&0,\label{eat}\\
\o^2 a^n_r-i\o\p_ra^n_0-\frac{in}{r^2}\p_ra^n_\t -\frac{n^2}{r^2}a^n_r- m_W^2(a^n_r-\p_r\sigma) &=&0\label{ear},
\eea
with $ m_W^2=e^2v^2$ the $W$ mass.
For the continuity of the fields we require that $a^n_\t$, $a^n_r$ vanish at the origin for all $n$, except for $a^0_0$ which might be constant.
Note that these four equations are not independent in the general case. A possible resolution is to work in the unitary gauge ($\sigma=0$). We extract $a^n_0$ from the equation (\ref{es}) and replace it in the equations (\ref{eat}, \ref{ear}):
\bea
\left(-\frac{r}{\D}\p_r\frac{\D}{r}\p_r-\omega^2+\frac{n^2}{r^2}+m_W^2\right)a^n_\theta-\frac{2in}{r}a^n_r&=&0\label{eAt},\\
\left(-\p_r\frac{1}{r\D}\p_r r\D-\omega^2+\frac{n^2}{r^2}+m_W^2\right)a^n_r+\frac{2in}{r^3}a^n_\theta&=&0.\label{eAr}
\eea
The equation (\ref{ea0}) for $a^n_0$ is not independent and can be dropped.
The analysis is slightly subtle and we have to treat all possible cases separately. 

Consider first $n=0$, the equations (\ref{eAt}, \ref{eAr}) decouple and read (for $r\leq R$)
\bea
\left(w^2-m_W^2\right) a^n_{\theta }(r)+\left(\frac{2 Mr-1}{r}\right) (a^n_{\theta })'(r)+(a^n_{\theta })''(r)&=&0,\label{0At}\\
\left(-m_W^2+w^2-\frac{1}{r^2}\right) a^n_r(r)+\left(\frac{2 Mr+1}{r}\right) (a^n_r)'(r)&&\\+(a^n_r)''(r)-2M \delta(r-R) a^n_r &=&0.\label{0Ar}
\eea
The presence of the $\delta$-function in the equation (\ref{0Ar}) changes the boundary condition on the brane for the symmetric mode to $(a^n_r)'(R-\eps)=-M a^n_r(R)=-(a^n_r)'(R+\eps)$, which makes the lowest energy mode for $a^n_r$ heavy ($\omega>M$).
For $a^n_\t$, we have the general solution
\beq
a^n_\t\propto e^{-r \left(M+\Omega_W\right)} r^2 L_{\frac{1}{2} \left(\frac{M}{\Omega_W}-3\right)}^2\left(2 r
   \Omega_W\right),
\eeq
with $\Omega_W=\sqrt{M^2+m_W^2-\omega^2}$. The lowest energy mode satisfy $(a^n_\t)'(R)=0$ and is light, $\omega_0^0=m_W+O(e^{-2MR})$ and the next mode has energy of the order $M$.

We shall now consider the case $n\neq 0$. 
Equations (\ref{eAt}, \ref{eAr}) are coupled and should in general be solved numerically. However if we notice that the mode we are looking for lives on the brane and therefore should have  
\beq F_{r\t}=-i n a_r^n(r)+\partial_r a_\theta^n(r)=0\label{F0},\eeq
 the equations (\ref{eAt}, \ref{eAr}) reduce to (for $r\leq R$)
\beq 
\left(-m_W^2+w^2-\frac{n^2}{r^2}\right) a^n_\t(r)+\left(2 M+\frac{1}{r}\right) (a^n_\t)'(r)+(a^n_\t)''(r)=0.
\eeq
This is similar to the equation for the Higgs mode, with $m_H$ replaced by $m_W$.
The modes are therefore similar to the Higgs ones, 
\beq
a^n_\t \propto e^{-r \left(M+\Omega_W\right)} r^n L_{-\frac{M}{2 \Omega_W}-n-\frac{1}{2}}^{2 n}\left(2
   r \Omega_W\right)
\eeq
and $a^n_r=-\frac{i}{n}\partial_r a_\theta^n(r)$ have to be continuous at the poles and one the brane. The lowest energy mode satisfy $(a^n_\t)'(R)=a^n_r(R)=0$ and has energy $\omega_0^n=\sqrt{m_W^2+\frac{n^2}{r^2}}+\mO(e^{-2MR})$. As for the Higgs, the next mode has energy $\omega_1>M$.
Like for $n=0$, no other low energy mode is found in the equations (\ref{eAt}, \ref{eAr}).

To summarise, up to corrections $\mO(e^{-2MR})\ll 1$ the low energy spectrum and the dispersion relations exactly match the usual 1+1 dimensional Abelian Higgs model.

\subsection{Effective action}

We can now built the effective 1+1 dimensional action for the Kaluza-Klein (KK) modes. To simplify the algebra, we will remove all the $\mO(e^{-2MR})$ suppressed terms.
We work in the unitary gauge and use the following decomposition for the fields:
\bea
\phi&=&v+\frac{1}{\sqrt{2\pi}}\sum_{k=0}^\infty\sum_{n=-\infty}^\infty e^{in\theta}h_k^n(r) H^n_k(t),\notag\\
A_\t&=&\frac{1}{\sqrt{2\pi}}\sum_{k=0}^\infty\sum_{n=-\infty}^\infty e^{in\theta}a_{\t,k}^n(r) A^n_k(t),\label{kkdec}\\
A_r&=&\frac{1}{\sqrt{2\pi}}\sum_{k=0}^\infty\sum_{n=-\infty}^\infty e^{in\theta}a_{r,k}^n(r) A^n_k(t),\notag\\
A_0&=&\frac{1}{\sqrt{2\pi}}\sum_{k=0}^\infty\sum_{n=-\infty}^\infty e^{in\theta}a_{0,k}^n(r) A^n_k(t),\notag
\eea
with $k$ labelling different KK modes, and $n$ labelling the angular momentum. Note that we have to impose that the fields are real, that is to say $(A^n)^*=A^{-n},~ (H^n)^*=H^{-n}$ and $(a_M^n)^*=a_M^{-n}, ~(h^n)^*=h^{-n}$. 

To get the 1+1 dimensional action for the whole tower of the KK modes we substitute the KK expansion (\ref{kkdec}, \ref{aa0}) in the Lagrangian (\ref{LH}), and integrate over the extra-dimension coordinate $r$.  In the following, we will consider only the $k=0$ mode in the KK expansion (\ref{kkdec}) to get the low energy effective theory. We will also neglect the back-reaction of heavy Kaluza-Klein modes on the low energy action, supposing that these corrections are suppressed. This assumption will be checked in the next section.


\paragraph{Gauge kinetic term}
Combining the relations (\ref{es}, \ref{aa0}) we get
\beq
a_0^n=\frac{\o^2_n-m_W^2}{n\o_n}a_\theta^n(r). \label{aa0}
\eeq
The expansion (\ref{kkdec}) is inserted in the gauge kinetic term
\beq
S_{GF}=\int dr~d\theta~ dt\bD\frac{1}{2}  
\left[ \frac{1}{r} F^2_{0\theta} + rF^2_{0r} - \frac{1}{r} F^2_{\theta r} \right].
\eeq
We use the formulas (\ref{aa0}, \ref{F0}), separate the $t$ and $r$ dependencies and get after some straightforward algebra (keeping only the $k=0$ mode)
\beq
S_{GF}=I \int\;dt\sum_n\frac{1}{2}\left(|\dot A^n|^2-\frac{(\o_n^2-m_W^2)^2}{\o_n^2}|A^n|^2\right),
\eeq
with 
\beq
I= \left[\int dr\frac{\D}{r}\left(|a_\t^n|^2-\frac{r^2}{n^2}|(a_{\t}^{n})'(r)|^2\right)\right]\label{ncat},
\eeq
the dot means derivative with respect to time and the prime derivative with respect to $r$. We fix the normalisation of $a_\t^n$ such that $I=1$
which, with the dispersion relation $\o^2_n=m_W^2+\frac{n^2}{R^2}$, leads to
\beq
S_{GF}=\int dt \sum_n\half\left(|\dot A^n|^2-\frac{(n^2/R^2)^2}{m_W^2+n^2/R^2}|A^n|^2\right).
\eeq

\paragraph{Higgs kinetic term}
We shall now consider the Higgs kinetic part of the action, 
\beq
S_{D}=\int dr~d\theta~ dt\bD\half \left[r|D_0\phi|^2-r|D_r\phi|^2-\frac{1}{r}|D_\t\phi|^2\right].
\eeq
Again, we insert the KK expansion (\ref{kkdec}), use the relations (\ref{aa0}, \ref{F0}) and get
\beq
S_{D}=\int dt \sum_n \left(\half m_W^2|A^n|^2 I_a+\half |\dot H^n|^2\int dr |h^n(r)|^2-\frac{n^2}{2R^2}|\dot H^n|^2 I_h\right)+S_D^{int},
\eeq
with 
\bea
I_a&=&\int \frac{\D}{r}\left[-r^2\frac{(\o_n^2-m_W^2)^2}{n^2\o_n^2}|a_\t^n|^2+\frac{r^2}{n^2}|(a_\t^n)'(r)|^2-|a_\t^n|^2\right]\label{Ia},\\
I_h&=&\frac{R^2}{n^2}\int dr\D r \left( |(h^n)'(r)|^2+ \frac{n^2}{2r^2}|h^n(r)|^2\right)\label{Ih}
\eea
and we normalise $h^n$ such that 
\beq
\int dr |h^n(r)|^2=1.\label{nch}
\eeq
The interactions between gauge field and Higgs read
\beq
S_D^{int}=\int dt\left(-\frac{e^2}{2\pi}\sum_{n,k,m} A^n A^m H^k H^{-n-m-k} I_b-\frac{e^2 v}{\sqrt{2\pi}}\sum_{n,m} A^n A^m H^{-n-m} I_c\right),
\eeq
with
\bea
I_b&=& \int \frac{r \D}{2} h^k h^{-n-m-k}\left(-\frac{n}{R^2\o_n}\frac{m}{R^2\o_m}a_\t^na_\t^m+ \frac{1}{nm}(a_\t^n)'(a_\t^m)'+\frac{1}{r^2}a_\t^na_\t^m\right),\label{Ib}\\
I_c&=& \int \frac{r \D}{2} h^{-n-m}\left(-\frac{n}{R^2\o_n}\frac{m}{R^2\o_m}a_\t^na_\t^m+\frac{1}{nm}(a_\t^n)'(a_\t^m)' +\frac{1}{r^2}a_\t^na_\t^m\right).\label{Ic}
\eea

\paragraph{Higgs potential}
The remaining  Higgs potential is
\beq
S_V=\int dr~d\theta~ dt\left[-r\bD\frac{\lambda}{2}\left(|\phi|^2-v^2\right)^2\right],
\eeq
which is expanded to
\bea
S_V&=&\int dt \left(-\frac{m_H^2}{2} \sum_n |H^n|^2-\frac{\lambda v}{\sqrt{2\pi}} \sum_{n,m}\notag
 H^n H^m H^{-n-m} I_d\right.\\ &&\left.-\frac{\lambda}{8\pi}\sum_{n,m,k} \label{svh}
H^n H^m H^k H^{-n-m-k}I_e\right),
\eea
with
\beq
I_d=\int \D r h^n h^m h^{-n-m},\quad I_e=\int \D r h^n h^m h^k h^{-n-m-k}. \label{Ide}
\eeq

\paragraph{Low energy theory}
In Appendix \ref{B} we compute that 
\beq I_a=-1-\frac{n^2}{R^2\o^2},\quad I_h=1\eeq
and to $\mO((MR)^{-1},(m_HR)^{-2})$, we have
\beq
I_b=\frac{M}{R},\quad
I_c=\sqrt{\frac{M}{R}},\quad I_d=\sqrt{\frac{M}{R}}, \quad I_e=\frac{M}{R},
\eeq
which gives us
\bea
S_{eff}&=&\int dt \sum_n \half \left[|\dot A^n_0(t)|^2-\left(m_W^2+\frac{n^2}{R^2}\right)|A^n_0(t)|^2 \right.\\&&\left.+|\dot H^n_0(t)|^2-\left(m_H^2+\frac{n^2}{R^2}\right)|H^n_0(t)|^2\right]+S^{int}_{eff}.
\eea
and up to $\mO((MR)^{-1},(m_HR)^{-2})$, 
\bea
S^{int}_{eff}&=&\int dt \left[\sum_{m,n}\left(-\frac{\lambda v\sqrt{M}}{\sqrt{2\pi R}}H^n H^m H^{-n-m}-\frac{e^2 v\sqrt{M}}{\sqrt{2\pi R}} A^nA^mH^{-n-m}\right)\right.\notag\\&&+\left.\sum_{m,n,k}\left(-\frac{\lambda M}{8\pi R}H^n H^m H^k H^{-n-m-k}-\frac{e^2 M}{4\pi R}A^nA^mH^kH^{-m-n-k}\right)\right].\label{Sinteff}
\eea
Comparing to (\ref{SH11}, \ref{9}), we see that up to at least $\mO((MR)^{-1},(m_HR)^{-2})$, we get the same interactions as in the 1+1 dimensional Abelian Higgs with the identifications 
\beq
\tilde{e}^2=e^2M, \quad \tilde{v}^2=v^2/M,\quad \tilde{\lambda}=\lambda M.
\eeq
Note that these identifications are just what we would expect from dimensional analysis.
We conclude that the theory (\ref{LH}) successfully reproduces the 1+1 dimensional Higgs model on the boundary of the Disk as its low energy limit.

\subsection{Validity of the Kaluza-Klein expansion}

Our computations are purely classical, and up to now we haven't taken into account the corrections to the effective action coming from interactions with heavy modes. 
This goes beyond the scope of this article and we will only describe here a region of the coupling parameter space where our semi-classical computations can be trusted. 


The first problem that occurs in most warped scenarios, is that the couplings become large far from the brane. For instance, this can be seen if we rescale the gauge field like $eA_M \to A_M$. The gauge kinetic term becomes $\frac{\Delta}{e^2} F_{MN}F^{MN}$ and we see that the effective charge is $\frac{e}{\sqrt{\Delta}}$. As $\Delta(r)$ becomes very small far from the brane, the effective charge becomes very large. This leads to large quantum corrections and large effective couplings between modes which probe the region with small $\Delta$. This problem first concerns very energetic modes that live in the bulk and can probe the region with small $\Delta$ but having strongly coupled heavy modes may also lead to large corrections to the low energy interactions.

\paragraph{Behaviour of the theory at very high energies}

We will compute here the effective interactions between heavy modes and show under which conditions interactions between light and heavy modes do not affect the low energy theory.
We only show explicit calculations for the Higgs modes. The gauge fields case is similar.

If we compute the effective theory for all the KK modes we get the terms in the action of the form (\ref{svh}, \ref{Ide})
$$-\frac{\tilde{\lambda}_{k_1,k_2,k_3,k_4}^{n,m,p,-n-m-p}}{8\pi R}H_{k_1}^n H_{k_2}^m H_{k_3}^p H^{-n-m-p}_{k_4}.$$
The quartic coupling between KK modes is given by
\beq
\tilde{\lambda}_{k_1,k_2,k_3,k_4}^{n_1,n_2,n_3,n_4}=\lambda \int r \D \left(\prod_{i=1}^{4} h_{k_i}^{n_i}\left(r\right)\right)dr .
\eeq
For KK modes with very high energy $\o^n_k\gg M$, this integral can be approximated as (see Appendix \ref{C})
\beq
\tilde{\lambda}_{k_1,k_2,k_3,k_4}^{n_1,n_2,n_3,n_4}\sim\frac{\lambda}{R}e^{2MR}\frac{\prod_{i=1}^{4}\sqrt{\o_{k_i}^{n_i}}}{\sum_{i=1}^{4}(\o_{k_i}^{n_i})^2}.\label{71}
\eeq
It can be shown numerically that this formula fits well for all $k_i\geq 1$. We shall now study what are the consequences on these large couplings.

\paragraph{Corrections to the nonrenormalisable couplings}
The simplest correction to the effective action is the appearance of a nonrenormalisable $\lambda^l_6 H_l^6$ coupling, which comes at tree level from the following diagram.
\begin{center}
\begin{pspicture}(0,-1.2)(6,1.3)
\rput(0.0,-1){$H_l$}
\rput(0.0,0.2){$H_l$}
\rput(0.0,1){$H_l$}
\rput(5.5,-1){$H_l$}
\rput(5.5,1){$H_l$}
\rput(5.5,-0.2){$H_l$}
\psline[linewidth=1pt,linestyle=solid](0.5,1)(2,0) 
\psline[linewidth=1pt,linestyle=solid](0.5,0.2)(2,0)
\psline[linewidth=1pt,linestyle=solid](0.5,-1)(2,0)
\psline[linewidth=1pt,linestyle=solid](5,1)(3.5,0)
\psline[linewidth=1pt,linestyle=solid](5,-1)(3.5,0)
\psline[linewidth=1pt,linestyle=solid](5,-0.2)(3.5,0)
\psline[linewidth=3pt,linestyle=solid](1.95,0)(3.55,0)
\rput(2.7,-0.5){$H_k$} 
\end{pspicture}
\end{center}
The contribution from heavy modes will be of the order
\beq
\tilde{\lambda}^l_6\sim\sum_k\tilde{\lambda}_{l,l,l,k}^2\frac{1}{\o_k^2}.
\eeq
For heavy KK modes, $l\geq 1$, we can use equations (\ref{71}) and the relation $\omega_k\sim \frac{\pi k}{2R}$, which leads to
\beq
\tilde{\lambda}^l_6\sim\lambda^2 e^{4MR} \sum_k\frac{l^3}{k(k^2+3l^2)^2}.
\eeq
The sum converges fast but the factor $e^{4MR}$ is very large. If we want our classical calculations to be valid, we have to impose a strong restriction on the 2+1 dimensional couplings $\lambda, e^2$, so that $\tilde{\lambda}_6$ and other corrections are small, that is
\beq
\lambda, e^2\ll M e^{-2MR}. \label{weakcoupl}
\eeq
This assumption will be considered in the following; it enables to perform semi-classical computations. Although the assumption (\ref{weakcoupl}) seems very restrictive, there is no obvious reason that the further results are only  valid under this constraint. We suppose that they are also valid under more general assumptions, but we will not prove it here. The extension to larger couplings requires lattice computations \cite{Laine:2004ji}. 

\section{Nonlinear solutions}

After discussing the perturbative spectrum of the theory we study solutions of the nonlinear equations of motion.
In the usual 2+1 dimensional Abelian Higgs model --- not localised on a brane --- neither sphaleron nor instanton exists. There only exists solitons --- Nielsen-Olesen vortices in two spatial dimensions.  The vortices possess the property that, if we consider a closed curve $C$ surrounding the vortex centre, the following relation is satisfied 
\beq \frac{e}{2\pi}\oint_C {\bf A}\cdot {\bf dl}=n,\label{topcharge}\eeq
where $n$ is an integer called topological charge of the vortex.

Processes discussed in \cite{Khlebnikov:2004am,Niemi:1983rq} can happen: starting from the three dimensional vacuum,  a vortex-antivortex pair is created near the south pole. The vortex can move across the equator up to the north pole.
The bosonic configuration of the final state looks on the equator (brane) like the $n$ vacua. Indeed using the relation (\ref{topcharge}), the Chern-Simons number is $$N_{CS}=\frac{e}{2\pi}\int dx A_1=n.$$
From the point of view of someone living on the brane, a topological transition (indistinguishable from an instanton transition) occurred and the vacua changed (from $|0\rangle$ to $|1\rangle$). The difference with the pure two dimensional theory is that the end point is not a \emph{vacuum} state.  It contains two vortices, which represent a large energy cost for the system.

We shall now see what happens when the warp factor is introduced and the fields are localised. The main effect of the warp factor is to damp the action of field configurations localised far from the brane. We therefore expect that, at the classical level, the vortex is light close to the pole and heavier when crossing the brane.

The 1+1 dimensional $|n\rangle$ vacua with $n\neq 0$ are not vacua any more in the higher dimensional theory. They become resonances with the energy of two vortices with topological charge $n$ and some suppressed decay rate.

\subsection{Vortex at the centre of the Disk}

With the warp factor, the vortex is repulsed from the brane and a stable solution of the equations of motion can only be found at the centre of the disk.  In this section we study the energy of the vortex at the centre of the disk and find its dependency on the parameters of the model and on its topological charge $n$. 
The vortex has the same properties as the 1+1 dimensional instanton that is to say $\phi(r=0)=0$, rotational symmetry and winding number $Q=n$. It is also parametrised by equations (\ref{phi}, \ref{AA}) but the functions $f(r),A(r)$ are deformed by the presence of the warp factor.
The energy of the $A(r),f(r)$ field configuration is
\bea
E=\pi v^2\mu^2\int dr\left\{\Dr A'^2+\bD\left[r f'^2+\frac{f^2}{r}(n-A)^2+r\frac{\mu^2}{2}(f^2-\mu^{-2})^2\right]\right\}
,\eea
with $\mu^2=\frac{\lambda}{e^2}$ and their equation of motion read
\bea
-\frac{r}{\bD}\p_r\D \frac{A'}{r}-f^2(n-A)&=&0,\notag\\
-\frac{r}{\bD}\p_r\bD r f'+h(n-A)^2+r^2\mu^2(f^2-\mu^{-2})f&=&0.\label{eomvortex}
\eea
In the following, we will solve numerically the equations (\ref{eomvortex}) with boundary conditions (\ref{l3}) as in \cite{Burnier:2005he}.
There are tree dimensionless parameters, one for the gauge-Higgs fields $\mu$, one for the warp factor $\bar M=\frac{M}{\sqrt{\lambda}v}$ and the size of the Disk $\bar R=\sqrt{\lambda}vR$. The dependencies of the energy with respect of $\bar M,\bar R$ and the topological charge $n$ will be studied. For the sake of simplicity we will set $\mu=1$.

\begin{figure} 
\begin{center} 
\includegraphics[width=100mm,height=70mm]{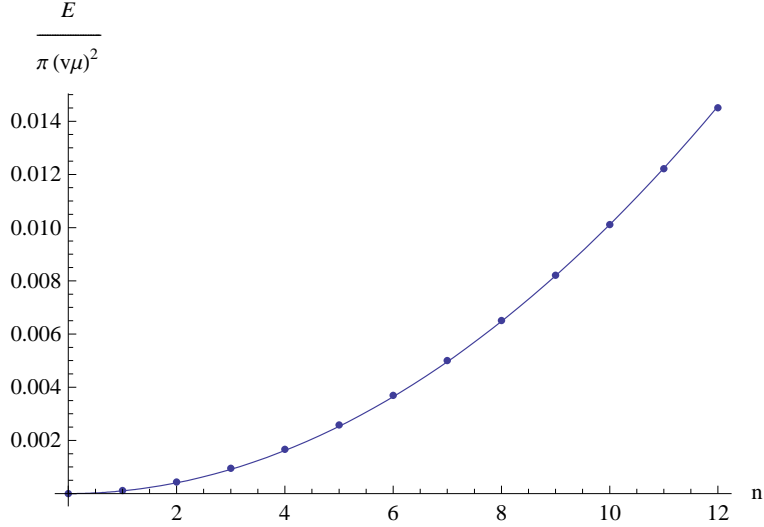} 
\caption{Vortex energy $\frac{E}{\pi v^2\mu^2}$ as function of the topological charge $n$, fitted with $0.000101 n^2$. Other parameters are fixed to $\bar{M}=2,~\bar{R}=3$.}
\label{E_n}
\end{center}
\end{figure}

The dependency on the topological charge is a parabola $E\propto n^2$ for fixed $\bar M,\bar R$. An example of this is shown in Fig. \ref{E_n}. We checked that for $\bar{M}=0..10$, $\bar{R}=1..6$, $n=0..10$, the discrepancy between the fit $E\propto n^2$ and the points is much less that $0.1\%$

\begin{figure} 
\begin{center} 
\includegraphics[width=100mm,height=70mm]{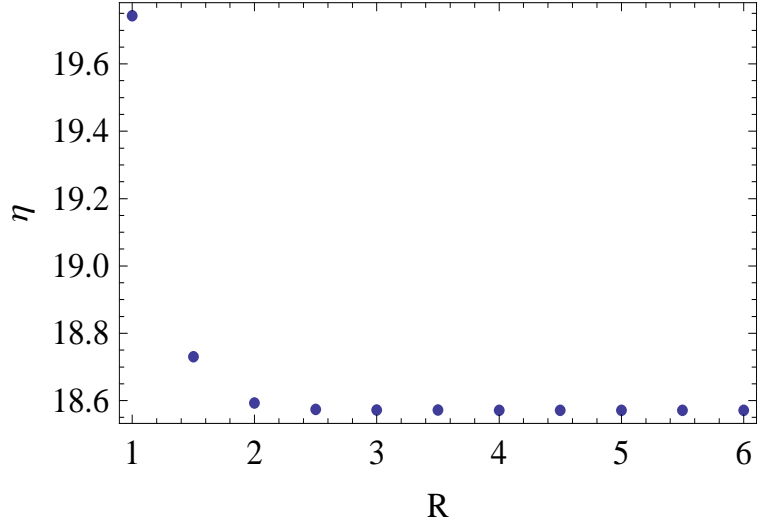} 
\caption{Parameter $\eta$ as function of the disk radius $\bar R$, Other parameters are $n=1$, $\bar{M}=4$}
\label{fE_R}
\end{center}
\end{figure}

The dependence on the radius size $\bar{R}$ is $E\propto \exp(-2\bar M\bar R)$, with some deviation from this behaviour at small values of $\bar{R}$, In Fig.\ \ref{fE_R}, we plot the value of 
\beq\eta=\frac{E}{\pi v^2\mu^2n^2\exp(-2\bar M\bar R)}.\label{eta}\eeq

\begin{figure} 
\begin{center} 
\includegraphics[width=100mm,height=70mm]{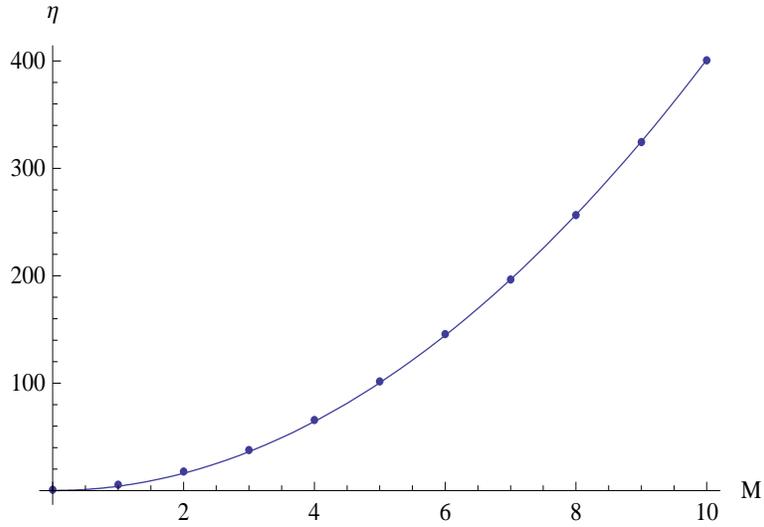} 
\caption{Parameter $\eta$ as function of the brane scale $\bar M$, fitted with $4.013 \bar M^2$. Discrepancies from points to the fits only occur for very small $\bar{M}$}
\label{fE_M}
\end{center}
\end{figure}

We will now extract the $\bar{M}$ dependency. We use of the previous dependencies and plot $\eta$ for $\bar{M}=0..10$. In Fig.\ \ref{fE_M}, $\eta$ is fitted with values of $n=0..9$, and the large $\bar{R}$ limit is used. For large $\bar{M}$ the dependency is again a parabola $E\propto \bar{M}^2$, and deviation from this occurs at small value of $\bar{M}$.

To sumarise, within the relevant region ($\bar{M}\gg 1\gg 1/\bar{R}$) we have the following vortex energy
\beq
E\simeq 4\pi v^2 \mu^2 \bar{M}^2 n^2 e^{-2\bar{M}\bar{R}}.
\eeq
Coming back to dimensionfull parameters, this takes the form
\beq
E\simeq 4\pi \frac{{M}^2}{e^2} n^2 e^{-2{M}{R}}.
\eeq

\subsection{Vortex sitting on the brane: Sphaleron}
%
The other case of interest is when a vortex of topological charge one is located on the brane, that is to say $\phi(R,\bar\t)=0$ at some $\t=\bar\t$; this is the 2+1 dimensional analog of the 1+1 dimensional sphaleron.
The presence of the brane breaks the rotational symmetry of the vortex. The exact solution is found by integrating numerically the 2 dimensional equations arising from the variation of the energy functional. Details of the numerical procedure are given in Appendix \ref{apD}.

\begin{figure} 
\begin{center} 
\includegraphics[width=100mm,height=70mm]{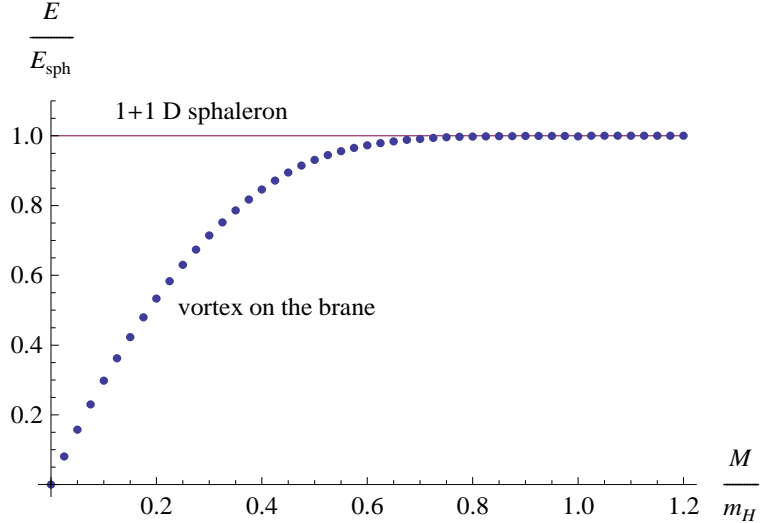} 
\caption{Energy of the different topological solutions as function of the brane mass-scale (for $\frac{m_H}{m_W}=1$). The energy is scaled so that the 1+1 dimensional sphaleron has energy 1. The energy of the vortex sitting on the brane meets the 1+1 dimensional sphaleron energy for large $M$. The accuracy of the results is estimated to be better than 0.1\%} 
\label{f2} 
\end{center} 
\end{figure} 

The results of the numerical simulation are shown in Fig.\ \ref{f2}. We only consider the case $m_H R\gg 1$. As long as $\bar{M}\gtrsim1$, the 2+1 dimensional sphaleron energy exactly matches the 1+1 dimensional one. Note that if the KK mass is larger than the sphaleron negative mode $\omega_-\sim 0.8 m_H$, the properties of the sphaleron are expected to be part of the low energy theory and match the 1+1 dimensional result (\ref{Esph}).

\subsection{Movement of the global mode in the potential - solution to the strong CP problem}

\begin{figure} 
\begin{center} 
\includegraphics[width=100mm,height=70mm]{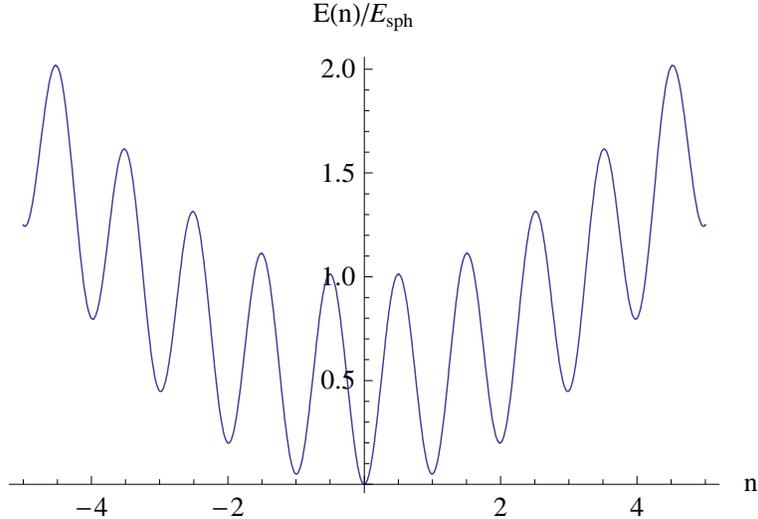} 
\caption{Sketch of the minimal energy for a configuration with winding number $n$. The value in local minima correspond to the energy of two vortices at the centre of the disk and the maxima contains the contribution of the sphaleron energy.}
\label{potq}
\end{center}
\end{figure}

The topological charge can be seen as a field \cite{Luscher:1978rn}, in the 1+1 dimensional Abelian Higgs model it has a periodic potential (see Fig.\ \ref{pot2D}). The addition of a compact extra-dimension breaks this periodicity (see Fig.\ \ref{potq}).

We shall compare the different scales in our model. As usual we consider the inequality $M\gg m_H\sim m_W\gg 1/R$ together with the 1+1 dimensional weak coupling requirement $\tilde{v}\gg1$ and the $2+1$ one (\ref{weakcoupl}). 
The sphaleron mass
\beq
\frac{m_{sph}}{m_H}\approx \tilde{v}^2\gg 1 \label{c1}
\eeq
is much larger that the particle mass in the weak coupling regime and much larger than the vortex mass,\footnote{In fact, the full set of requirements for the parameters is the following $\frac{m_H^2}{v^2M}
\ll
e^{-2MR}
\ll
\frac{m_H^3}{M^3}
\ll
1$.  The first inequality is the weak coupling of the 2+1 dimensional theory, the second is (\ref{c2}) and the third says that the KK modes are heavier, then the localised ones.}
\beq
\frac{m_{sph}}{m_{vortex}}\approx\frac{m_H m_w^2}{M^3 e^{-2MR}}\gg 1. \label{c2}
\eeq
At first sight, the energy of the vortex $E_v(n)$, and therefore the energy $E_n=2E_v(n)$ of the state $|n\rangle$ seems very small. However if we consider that the couplings (electric charge here) have to be tiny to be in the regime where our classical approximations are valid, we have\footnote{In fact, even stronger inequality is true, $m_{vortex}\gg M$.}
\beq
\frac{m_{vortex}}{m_H}=\frac{M^2}{e^2 m_H e^{2MR}}\gg \frac{M}{m_H}\gg 1. \label{c3}
\eeq
The latter relation shows that vortex pairs can decay to particles and the system relaxes to the $n=0$ vacuum. In this case the strong CP problem is solved. Indeed the 1+1 dimensional instanton transitions which make the average of the topological density ($\langle q\rangle\neq0$) non-vanishing are not anymore instanton transitions and can occur only if a large energy is available in the system. At low energy, for example for the measurement of the neutron dipole moment, a topological transition cannot occur, the topological density vanishes ($\langle q\rangle=0$ since already $\langle |q|\rangle=0$) therefore $\theta$ is effectively zero.

In the strong coupling regime, if the mass of the vortex happens to be too small the states $|n\rangle$ will be almost degenerate and we expect that the strong CP problem reappear at some point. Indeed when the $|n\rangle$ are almost degenerate, the real vacuum will be a supperposition of several $|n\rangle$. Several different supperposition of these $|n\rangle$ states with different phases and slightly higher energy will exist and  we expect to have some sort of discrete strong CP problem there.  The precise analysis of these states is left for future work.

\section{Conclusion}

Extra-dimensions might give an appealing solution to the strong CP problem. The solution relies on a global field dynamic and does not imply the existence of a new particle. Unlike for the axion solution, it cannot be ruled out by astrophysical observations. However, the processes with change of the topological number (and, at the same time, fermion number) proceed differently in the proposed setup. 

In this paper the 2+1 dimensional toy model was studied, which at low energy reduces to the 1+1 Abelian Higgs model, often used to model some QCD properties.  If the inequalities $m_{sph}\gg m_{vortex}\gg m_H, m_W$ are satisfied, the 1+1 dimensional $|n\rangle$ vacua become, for $n\neq 0$, resonances in the 2+1 dimensional theory. These resonances have a high energy and decay reasonably fast to particles. The system relaxes to the $|n=0\rangle$ vacuum, which is the true vacuum of the 2+1 dimensional theory. The vacuum structure is trivial and the strong CP problem is solved in this case. We showed that these inequalities (\ref{c1}, \ref{c2}, \ref{c3}) are realised in the weak coupling regime (\ref{weakcoupl}). We suppose that these relations will remain true in the strong coupling regime. This seems reasonable for the relations (\ref{c1}, \ref{c2}) while for (\ref{c3}), it seems to fail, since we used the weak coupling assumption. However in the strong coupling case, the quantum corrections are large and, unlike the classical action, are not suppressed by the warp factor at the vortex centre. Indeed quantum corrections are expected to be of order $\hbar$ independently of the normalisation of the action. It would be interesting to study the strong coupling regime with lattice simulations to check whether this large extra-dimension solution to the strong CP problem also works at large couplings.

We choose in this paper a very simple geometry, which allows for simple analysis but which has the annoying property that the size of the universe is just $2\pi$ larger than the extra-dimension. This restricts the semi-classical calculations to very weak couplings. A way to avoid strong coupling problems would be to consider a more complicated geometry, which allows for unrelated universe and extra-dimension size. We could then have a very large or infinite space and a small enough extra-dimension. The set-up will be more complicated and may not enable a simple analysis of the nonlinear solutions. However as we have shown here, it seems that the localisation of the fields on a brane can be done independently of the bulk geometry. It is indeed an interesting result on its own that an Abelian Higgs can be localised on the boundary of a disk.

Finally, we may speculate how the obtained results transfer to the real world QCD. We expect that the degeneracy of the $n$-vacua will be lifted by the extra-dimension and the new $|n\rangle$ states (or resonances) will have an energy monotonically increasing with $n$. The tunnelling between the $n$ and $n-1$ will still be suppressed by the instanton (or, strictly speaking, bounce) action. At very high energy, these bounce transitions are possible and will look like normal instanton transitions. If we can arrange the geometry so that the states (or resonances) $|n\neq 0\rangle$ have sufficiently large energy, bounces will not contribute in a measurement of the neutron dipole moment. The topological charge density will be zero and there will not be any observable effect of a $\theta$-angle, so that the strong CP problem will be solved.


\acknowledgments

The authors thank S. Khlebnikov, C. Becker and M. Laine for helpful discussions and F. K\"uhnel for rereading the manuscript. We are especially grateful to M. Shaposhnikov for valuable discussions at all stages of the
work.

\appendix

\section{Spectrum of Kaluza-Klein modes}\label{A}

\paragraph{Higgs dispersion relation}
We derive the dispersion relation for the light mode of the Higgs. To this aim, we have to impose the boundary condition on the brane, $h'(R)=0$. Suppose that the low energy mode satisfy $\omega\sim m_H\ll M$. We can then use the asymptotic expansion for large argument of Laguerre function :
\beq
L_\nu^\lambda(z) \overset{z\to\infty}{\longrightarrow} \sin(\nu \pi)\Gamma(\lambda+\nu+1)z^{-\lambda-\nu-1}e^z\left(1+\frac{(1+\nu)(1+\lambda+\nu)}{z}+\mO(z^{-2})\right).\label{Linf}
\eeq
In this limit, equation (\ref{hsg}) becomes
\beq
h_0^n(r)\propto e^{-\frac{\dw^2r}{2 M}}(2 r M)^{-\frac{\dw^2}{4 M^2}}\left(1-\frac{n^2}{2 M r}+\mO(M^{-2})\right),\label{hinf}
\eeq
with $\dw^2=\omega^2-m_H^2$. Imposing $h_0'(R)=0$ gives $\dw^2=\frac{n^2}{R^2}$.

\paragraph{Gauge field dispersion relation}
The computation is the same as for the Higgs, the function $a_\t^n(r)$ being the same as $h_0^n(r)$ replacing $m_H$ by $m_W$. Imposing $a_\t'(R)=0$ implies $\o^2 =m_w^2+\frac{n^2}{R^2}$. Note that, form equations (\ref{aa0}, \ref{F0}) this implies $a_r^n(R)=0$ and $a_0'(R)=0$, which are the correct boundary conditions for these fields.

\paragraph{High energy modes for the Higgs}

Using the expansion (\ref{Linf}), we see that for $\omega\gg M$, 
\beq h^n(r)\sim e^{i r \omega}.\eeq
One of the requirement $h(R)=0$ or $h'(R)=0$ are satisfied if $R\omega= k\pi/2$, which gives $\omega_k-\omega_{k-1}=\frac{\pi}{2R}$.

\section{Couplings in the effective action} \label{B}

In this section, we compute the parameters $I$. We compute exactly the parameters related to the quadratic part of the action, first $I_a$. We start from the normalisation condition for $a_\t^n$ (\ref{ncat}), integrate $|(a_\t^n)'(r)|^2$ by part, use the equation of motion (\ref{eAt}) for $a_\t^n$ and the relation (\ref{F0}), 
\beas
1&=&\int dr \frac{\D}{r}\left(a_\theta^n a_\theta^{-n}-\frac{r^2}{n^2}(a_\t^n)'(r)(a_\t^{-n})'(r)\right)\\
&=&\frac{\o^2_n-m_W^2}{n^2} \int dr r\D |a^n_\t(r)|^2.
\eeas
From which we get 
\beq
\int dr r\D |a^n_\t(r)|^2=\frac{n^2}{\o^2_n-m_W^2}.
\eeq
This, together with the normalisation condition (\ref{ncat}) gives us
\beq 
I_a=-1-\frac{n^2}{\o^2_n-m_W^2}.\frac{(\o^2_n-m_W^2)^2}{n^2\o^2_n}=-1-\frac{n^2}{R^2\o_n^2}.
\eeq
To calculate $I_h$, we integrate $|(h^n)'|^2$ by part and use the equation of motion (\ref{equah})
\beas
I_h&=&\frac{R^2}{n^2}\int dr r\D\left( (h^n)'(r)(h^{-n})'(r)+ \frac{n^2}{2r^2}|h^n(r)|^2\right)\\
&=&\frac{R^2}{n^2}\int dr r\D (\o^2_n-m_H^2)|h^n(r)|^2,
\eeas
which, using the normalisation condition (\ref{nch}) and the dispersion relation, gives $I_h=1$.

The other parameters $I_b,I_c,I_d,I_e$ are computed to $\mO((MR)^{-1},(m_HR)^{-2})$ using the asymptotic expansion (\ref{Linf}) for the mode functions. Indeed because of the presence of the warp-factor and the fact that the modes $a_\t^n, h^n$ are roughly constant, the integrals $I_{b,c,d,e}$ are dominated by values close to the brane $(r=R)$. At these points, the argument of the Laguerre functions is large and the asymptotic expansion can be used. We first compute $I_d, I_e$ (\ref{Ide}), to this aim, we first have to normalise $h^n(r)$. From (\ref{hinf}), and imposing $\int dr r\D|h^n(r)|^2=1$, we get
\bea
h^n&=& e^{M(R-r)-r\Omega_H}\frac{ r^n (2\Omega)^{n+\half-\frac{M}{2\Omega_H}}}{\sqrt{2R} ~\Gamma\left(n+\half-\frac{M}{2\Omega_H}\right)}L^{2n}_{-n-\half-\frac{M}{2\Omega_H}}(2r\Omega_H)\label{hinfn}
\\&\overset{\Omega_H r\gg 1}{\longrightarrow}&
\sqrt{\frac{M}{R}}\left(\frac{R}{r}\right)^{-\frac{n^2}{4R^2M^2}}e^{-(r-R)\frac{n^2}{2MR^2}}\left(1+\mO\left(\frac{1}{MR}\right)\right)\notag .
\eea
Some straightforward algebra gives
\bea
I_d&=&\int dr~ \D r~h^n h^m h^{-n-m}=\sqrt{\frac{M}{R}}\left(1+\mO\left(\frac{1}{MR}\right)\right),\\
I_e&=&\int dr~\D r ~h^n h^m h^k h^{-n-m-k}=\frac{M}{R}\left(1+\mO\left(\frac{1}{MR}\right)\right).
\eea
Note that these integrals, at least at leading order, do not depend on the values of $n,m,k$.

To compute the remaining $I_b,I_c$, we need the asymptotic expansion for $a_\t^n(r)$. It only differs from $h^n(r)$ by the normalisation condition. As $a_\t^n(r)$ is constant near $r=R$, we can neglect the term $|(a_\t^n)'(r)|^2$ in the normalisation condition (\ref{ncat}), which becomes $\int dr\frac{\D}{r}|a_\t^n|^2=1$. This implies that
\beq
a_\t^n=\sqrt{MR}\left(\frac{R}{r}\right)^{-\frac{n^2}{4R^2M^2}}e^{-(r-R)\frac{n^2}{2MR^2}}\left(1+\mO\left(\frac{1}{MR}\right)\right). \label{ainfn}
\eeq
Using (\ref{hinfn},\ref{ainfn}) and performing some straightforward algebra gives
\beas
I_b&=& \int dr~\frac{r \D}{2} h^k h^{-n-m-k}\left(-\frac{n}{R^2\o_n}\frac{m}{R^2\o_m}a_\t^na_\t^m+ \frac{1}{nm}(a_\t^n)'(a_\t^m)'+\frac{1}{r^2}a_\t^na_\t^m\right)\\
&=& \int dr~\frac{r \D}{2} h^k h^{-n-m-k}\frac{1}{r^2}a_\t^na_\t^m\left(1+\mO\left(\frac{1}{m_W^2R^2}\right)\right)\\
&=&\frac{M}{R}\left(1+\mO\left(\frac{1}{MR},\frac{1}{m_W^2R^2}\right)\right)
\eeas
and
\beas
I_c&=& \int dr~\frac{r \D}{2} h^{-n-m}\left(-\frac{n}{R^2\o_n}\frac{m}{R^2\o_m}a_\t^na_\t^m+\frac{1}{nm}(a_\t^n)'(a_\t^m)' +\frac{1}{r^2}a_\t^na_\t^m\right)\\
&=&\int dr~\frac{r \D}{2} h^{-n-m}\frac{1}{r^2}a_\t^na_\t^m\left(1+\mO\left(\frac{1}{m_W^2R^2}\right)\right)\\
&=&\sqrt{\frac{M}{R}}\left(1+\mO\left(\frac{1}{MR},\frac{1}{m_W^2R^2}\right)\right).
\eeas

\section{Couplings at very high energy}\label{C}

We already computed the effective coupling $\lambda=\lambda_{0,0,0,0}=M\tilde{\lambda}$ in the low energy effective action (\ref{Sinteff}) but we are now interested in couplings $\lambda_{k_1,k_2,k_3,k_4}$, with $k_i$ large.

At very high energy ($\o^k_n\gg M$), the wave functions $h^n_k(r)$ are large near the centre of the disk ($r=0$). The integrals for the effective couplings will be saturated by values close to $r=0$, therefore we make use of the small argument expansion of the Laguerre functions
\beq
L_\nu^\lambda(z)\overset{z\ll 1}{\longrightarrow}\frac{\Gamma(\nu+\lambda+1)}{\Gamma(\nu+1)}\left(\frac{1}{\Gamma(\lambda+1)}-\frac{\nu z}{\Gamma(\lambda+2)}+\dots\right).
\eeq
This expansion is introduced in the wave function $h^n_k$:
\beq
h^n_k(r)\overset{rM\ll 1}{\longrightarrow} e^{+M(R-r)-r(\Omega_H)} \frac{r^n(2\omega)^{n+\half-\frac{M}{2\Omega_H}}}{\sqrt{2R}\Gamma(2n+1)\Gamma\left(-n-\frac{M}{2\Omega_H}+\half\right)}.
\eeq
While computing the effective action, we will identify
\bea
\frac{\lambda_{k_1,k_2,k_3,k_4}}{8\pi R}&=&\lambda\int r \D dr \prod_{i=1}^4 h^{n_i}_{k_i}(r)\notag\\
&\approx& \frac{2\lambda e^{2MR}}{\pi^2 R^2}\frac{\prod_i\sqrt{\o_i}}{\sum_i \o_i^2}\gamma(n_1,n_2,n_3,n_4),
\eea
with $\gamma(0,0,0,0)=1$ and $\gamma(n_1,n_2,n_3,n_4)$ of order unity.

\section{Vortex sitting on the brane, numerical analysis}\label{apD}

To simplify slightly the analysis, we notice that the vortex solution has size of a few $1/m_H$ which is in our assumptions $M\gg m_H\gg 1/R$ much smaller that the radius of the disk. Furthermore what happens away from the brane is damped by the warp factor. We can therefore consider that the disk looks like a half plane in the region where the vortex lies. The vortex on the brane still possesses one discrete symmetry: the reflection along the perpendicular to the brane passing through its centre. We will therefore solve the equations on a quarter of plane.
To get a better resolution of the instanton centre, we parametrise the quarter of plane in polar coordinates $r\in ]0,\infty[,~\theta\in[0,\pi/2]$.

We choose the $A_0=0$ gauge and use the remaining time independent gauge freedom to cancel the radial component of the vector field\footnote{We are dealing with static (time independent) configuration}: $A_r=0$. We also use dimensionless variables as in (\ref{dimless}). The energy functional reads:
\bea
E[H,\sigma,A_\theta]&=
&v^2\mu^2\int rd r d \theta e^{-2 M r |\sin\theta|} \left(\frac{1}{2}(\partial_r A + \frac{A}{r})^2+\frac{1}{2}\left(\partial_rf\right)^2+\frac{1}{2r^2}\left(\partial_\theta f\right)^2\right.\notag \\&& \left.+\frac{f^2}{2}\left[\left(\partial_r\sigma\right)^2+\left(\frac{1}{r}\partial_\theta\sigma-A\right)^2\right]+\frac{\mu^2}{4}(f^2-\frac{1}{\mu^2})^2
\right).\label{Estat1}
\eea

Numerically, we minimise the energy functional using as starting configuration, the usual vortex configuration $A(r), f(r)$ from (\ref{phi},\ref{AA}).
\bea 
A(r,\theta)=A(r),\quad f(r,\theta)=f(r)
, \quad \sigma(r,\theta)=\theta.
\eea 
The convergence to the exact solution is achieved by introducing an artificial time dependency and solving Hamiltonian equations:
\bea \frac{d}{dt}A_\theta=-\frac{\delta E}{\delta A_\theta},\quad \frac{d}{dt}H=-\frac{\delta E}{\delta H},\quad \frac{d}{dt}\sigma=-\frac{\delta E}{\delta \sigma}.\eea
Note that an upper bound on the vortex energy can be found integrating the energy functional (\ref{Estat1}) with the usual vortex configuration in a flat 2+1 dimensional space (\ref{phi},\ref{AA}).

To avoid trivial divergences at $r=0$, we start the integration at $r=\eps\sim 10^{-5}$ and consider a finite radius $R$ to end the integration.\footnote{In dimensionless units we take $\bar R=5$. We checked that this do not introduce sizable errors.} As we use polar coordinates to describe the Higgs field, we should also arrange the boundary conditions to avoid that $H$ vanishes near $r=\eps$ and rescale the field $\sigma$ such that it is single valued at $r=0$, $\sigma(r,\theta)\to \sqrt{r}\sigma(r,\theta)$, this allows to fix correctly the value of $\sigma$ so that the action remain finite\footnote{The action near $r=0$ contains a term proportional to $r^3(\partial_r\sigma)^2$, which is finite if $\sigma\sim r^a$, with $a>-1/2$, that is to say $\sqrt{r}\sigma=\bar{\sigma}\to 0$ as $r\to 0$} at $r\to 0$.
Boundary conditions are set on the four edges; 
\bea
A_\theta(\eps,\theta)=a(\eps)\sim0,\quad H(\eps,\theta)=f(\eps)\sim0,\quad\sigma(\eps,\theta)=\theta\sqrt{\eps}\sim0,\\
\partial_\theta A_\theta(r,0)=0, \quad\partial_\theta H(r,0)=0, \quad\sigma(r,0)=0,\\
A_\theta(R,\theta)+R \partial_r A_\theta(R,\theta)=0,\quad H(R,\theta)=\frac{1}{\mu},\quad \sigma(r,\theta)-2 R \partial_r\sigma(R,\theta)=0,\\
\partial_\theta A_\theta(r,\frac{\pi}{2})=0, \quad\partial_\theta H(r,\frac{\pi}{2})=0, \quad\sigma(r,\frac{\pi}{2})=\frac{\pi}{2}\sqrt{r}.
\eea
The conditions at $r=0$ and $r=R$ are chosen to get a finite action and the conditions at $\theta=0$ and $\theta=\pi/2$ are dictated by the discrete symmetries and the topological charge.

\end{document}